\documentclass[12pt,a4paper]{article}
\usepackage{jheppub}  
\usepackage[small]{caption}
\usepackage{braket}
\def\een{\end{equation}}

\def\be{\begin{equation}}
\def\ee{\end{equation}}
\def\ba{\begin{array}}
\def\ea{\end{array}}

\def\dalemb#1#2{{\vbox{\hrule height .#2pt
        \hbox{\vrule width.#2pt height#1pt \kern#1pt
                \vrule width.#2pt}
        \hrule height.#2pt}}}

\newcommand{\bea}{\begin{eqnarray}}
\newcommand{\eea}{\end{eqnarray}}

\newcommand{\Tr}{{\rm Tr}\,}

\title{Large rank Wilson loops in ${\cal N}=2$ superconformal QCD at strong coupling }
\author{Benedict Fraser and S. Prem Kumar}
\affiliation{ 
 Department of Physics, \\
 Swansea University, \\ 
Singleton Park,\\
 Swansea, SA2 8PP, U.K.  
}
\emailAdd{s.p.kumar@swansea.ac.uk}
\emailAdd{pyfraser@swansea.ac.uk}
\abstract{We compute the expectation values of circular Wilson loops in large representations at strong coupling, in the large-$N$ limit of the ${\cal N}=2$ superconformal theory with $SU(N)$ gauge group and $2N$ hypermultiplets. Employing Pestun's matrix integral, we focus attention on symmetric and antisymmetric representations with ranks of order $N$. We find that large rank antisymmetric loops are independent of the coupling at strong 't Hooft coupling while symmetric Wilson loops grow exponentially with it. Symmetric loops display a non-analyticity  as a function of the rank, characterized by the splitting of a single matrix model eigenvalue from the continuum, bearing close resemblance to Bose-Einstein condensation in an ideal gas. We discuss implications of these  for a putative large-$N$ string dual. The method of calculation we adopt makes explicit the connection to Fermi and Bose gas descriptions and also suggests a tantalizing connection of the above system to a multichannel Kondo model.}
\begin{document}
\maketitle

\section{Introduction and summary}
Wilson loops, while being fundamental probes of gauge theory physics, following the development of the AdS/CFT correspondence
\cite{maldacena}, are now also known to play a key role as probes of string duals of large-$N$ gauge theories \cite{malwil, reyyee,dgo}. This aspect of Wilson loops was sharpened by the beautiful results of \cite{esz, dg} for supersymmetric BPS Wilson loops in ${\cal N}=4$ supersymmetric (SUSY) Yang-Mills theory. The upshot of these works was the conjecture that circular supersymmetric Wilson loops in ${\cal N}=4$ SYM are computed exactly, to all orders in the gauge  coupling, by the Gaussian matrix model.  This conjecture was subsequently proven by Pestun \cite{pestun} by considering 
${\cal N}=2$ and ${\cal N}=4$ SUSY gauge theories on a four-sphere and computing their partition functions  using powerful localization techniques. 

One of the theories for which Pestun provided a finite dimensional matrix integral representation is the $SU(N)$ ${\cal N}=2$ superconformal theory (SCFT) with $N_f=2 N$ flavour hypermultiplets (also referred to as the $A_1$ ${\cal N}=2$ SCFT). Recently, Passerini and Zarembo \cite{pz} explored properties of this matrix integral in the large-$N$ limit and deduced the behaviour of the circular Wilson loop in the fundamental representation, at strong 't Hooft coupling. The focus of the present paper is to use the large-$N$ matrix integral for the ${\cal N}=2$ SCFT to explore properties of supersymmetric Wilson loops in large representations, when the rank of the representation is of order $N$. Our primary motivation is to compare and contrast the results with  corresponding quantities in the ${\cal N}=4$ theory at strong coupling, and hence draw inferences about the possible nature of the large-$N$ string dual, if any, of the ${\cal N}=2$ SCFT.

The $N_f=2 N$ superconformal theory  has vanishing beta function and therefore an exactly marginal gauge coupling. Finding a string dual to this theory in the large-$N$ Veneziano limit \cite{veneziano} at strong 't Hooft coupling, is a long standing problem. Recent proposals in this direction include \cite{gpr, gaiottomal, bogdan, colgain}. 
In all cases the proposed backgrounds either contain regions of high curvature or are partly non-geometric as in \cite{gpr, colgain}. It is therefore interesting to ask if field theory probes such as Wilson loops can shed light on features of a putative large-$N$ string dual. Experience with the ${\cal N}=4$ theory indicates that Wilson loops in generic (large) representations can indeed act as effective probes of the dual geometry. In particular Wilson loops in symmetric and antisymmetric tensor representations are computed by probe D3 and D5-branes in $AdS_5\times S^5$ \cite{drukkerfiol, yamaguchi1, polloops, yamaguchi2, Gomis:2006sb, hep-th/0604209, mm,hep-th/0612022}. These are known to have world-volumes $AdS_2\times S^2 \subset AdS_5$ and $AdS_2 \times S^4$ ($S^4\subset S^5$) respectively, and thus are sensitive to different aspects of the geometry.

In \cite{pz}, a careful evaluation of the circular Wilson loop in the fundamental representation was performed at strong coupling in the ${\cal N}=2$ SCFT, making use of the corresponding matrix model. This led to a natural identification of a dual effective string tension  $T_{\rm eff} =\frac{3}{2\pi}\ln\lambda$, where $\lambda$ is the (large) 't Hooft coupling. The curious logarithmic dependence on the 't Hooft coupling translates to a non-exponential growth of the Wilson loop with $\lambda$ at large $\lambda$. The non-exponential dependence had already been deduced via scaling arguments applied to the associated matrix model in \cite{reysuyama}.
This is to be contrasted with the circular Wilson loop in ${\cal N}=4$ SYM which grows exponentially at strong coupling as $\sim e^{\sqrt\lambda}$. 

A particularly curious feature of the one-matrix model for the $A_1$ ${\cal N}=2$ SCFT is that the large-$N$ eigenvalue distribution
 has an infinite support at infinite 't Hooft coupling, with most of the eigenvalues remaining finite\footnote{This was realized in \cite{pz}. In earlier work \cite{reysuyama}, it was noted, using scaling arguments that the eigenvalue density approaches a limiting shape.}
. The limiting form of the eigenvalue distribution (or the large-$N$ ``master field'') does not describe the fundamental Wilson loop since the latter diverges in the limit $\lambda \to \infty$. However, in this paper we show that the limiting distribution does describe Wilson loops in large enough tensor representations. 
 
 Specifically, Wilson loops in the antisymmetric tensor representation with rank $k$ of order $N$ can converge to a result that is  independent of $\lambda$. In the large-$N$ limit with $f\equiv\frac{k}{N}$ fixed, we find that the antisymmetric Wilson loops are determined by the endpoint of the eigenvalue distribution only for $f\ll \sqrt{\ln\lambda}/\lambda\ll1$, and beyond this range approach a regular limit determined completely by the limiting eigenvalue distribution at infinite 't Hooft coupling. This behaviour strongly suggests that the internal/compact factors in the large-$N$ string dual, probed by the corresponding D-branes, remain highly curved or non-geometric as in \cite{gpr}.
 
 The situation with Wilson loops in the symmetric tensor representation of rank $k$ turns out to be somewhat different. Their expectation values are determined by the endpoint of the eigenvalue distribution at strong coupling and essentially track the behaviour of the fundamental Wilson loop up to a critical value of $f=f_c \sim \sqrt{\ln\lambda}/\lambda$. Beyond this point the Wilson loop experiences a non-analyticity characterized by the splitting of a single eigenvalue from the rest of the large-$N$ distribution. A related non-analyticity was observed for the symmetric loop in ${\cal N}=4$ SYM, in \cite{mm}. In that case, the position of the split eigenvalue
from  the large-$N$ distribution was mapped to the position of the  probe D3-brane in $AdS_5\times S^5$ \cite{yamaguchi1} which computes the symmetric Wilson loop.

The approach we use to compute the large rank Wilson loops is identical to that of \cite{mm}. In this paper we further emphasize the connection of the symmetric and anti-symmetric representations to the free Fermi and Bose gas pictures. This in turn suggests a potential tantalizing connection to the multichannel Kondo model \cite{parcollet, Sachdev:2010uz, muck, hkt}.

The most straightforward inference we can draw from our results is that the string dual to the ${\cal N}=2$ SCFT should have a weakly curved $AdS_5$ part which can be  probed by any D-branes that compute Wilson loops in symmetric representations.  The exponential growth of the latter with the 't Hooft coupling indicates that the corresponding D-branes  which compute them must be semiclassical, bearing some resemblance to the situation in the ${\cal N}=4$ theory. On the other hand, the behaviour of Wilson loops in the antisymmetric representation suggests that the internal or compact factor of the geometry must be highly curved. Our results may be viewed as predictions for the tensions of corresponding probe D-branes.

\section{The matrix model}

In \cite{pestun} it was shown that circular, supersymmetric Wilson loops in ${\cal N}=4$ and ${\cal N}=2$ superconformal gauge theories on $S^4$ are computed by matrix integrals as a consequence of localization of the partition function onto constant configurations. For an ${\cal N}=2$ SUSY theory, the supersymmetric Wilson loop, labelled by some representation $R$ of $SU(N)$, is defined as
\be
W_R ( C ) \,=\, {\rm Tr}_R\,\left[{\cal P}\exp \left(i\oint_C dt(A_\mu\,\dot x^\mu + i\Phi_I n^I |\dot x|)\right) \right],
\ee
where $\Phi_I$, $(I=1,2)$ are the two real adjoint scalars of the ${\cal N}=2$ vector multiplet and $n^I$ is a constant unit vector in $\mathbb R^2$. Any specific choice for this vector breaks the $SU(2)_R\times U(1)_R$ global R-symmetry of the ${\cal N}=2$ SCFT to $SU(2)_R$. In the $A_1$, ${\cal N}=2$ SCFT with $SU(N)$ gauge group and $N_f=2N$ hypermultiplets formulated on the four-sphere, the expectation value of the supersymmetric Wilson loop $W_R(C)$ is 
computed by a corresponding observable in a one-matrix model,
\be
\langle W_R ( C ) \rangle_{{\cal N}=2}\,=\,\langle {\rm Tr}_R e^ {2\pi\, a}\rangle_{\rm mm}\,.
\ee
The $N\times N$ matrix ``$a$'' can be viewed as the (constant) VEV for one of the adjoint scalars in the vector multiplet. In the diagonal basis for $a = {\rm diag}(a_1,a_2,\ldots a_N)$ with $\sum a_i=0$, the partition function on $S^4$ is obtained by integrating out all massive adjoint modes and the fundamental flavour fields at one-loop order, and subsequently integrating over the diagonal VEVs
\be
\langle W_R( C )\rangle\,=\,\frac{1}{{\rm Vol}(SU(N))}\int \,[d a]\,e^{-\frac{8\pi^2}{g^2}\,\Tr a^2}\,{\cal Z}_{\rm 1-loop}(a)
\,|{\cal Z}_{\rm inst}(a)|^2\, \Tr_{R}\,e^{2\pi a}\,.
\label{mm}
\ee
We have suppressed the explicit form of the integral in terms of the eigenvalues $\{a_i\}$. In particular, in the diagonal gauge the measure factor will give rise to a Vandermonde determinant. The first term in the integrand arises from the conformal coupling of the scalar fields to the curvature of $S^4$. The perturbative contributions are  1-loop exact obtained by integrating out fluctuations about the VEV, and ${\cal Z}_{\rm inst}$ is Nekrasov's instanton partition function \cite{nekrasov} for the equivariant theory on ${\mathbb R}^4$. In this paper we will be interested in the large-$N$ limit in which we expect instanton contributions to be exponentially suppressed and a careful consideration of these terms confirms this expectation \cite{pz}. In terms of the eigenvalues, the contribution from 1-loop fluctuations takes the form,
\be
{\cal Z}_{\rm 1-loop}\,=\,\frac{\prod_{i < j} H^2(a_i-a_j)}{\prod_i H^{2N}(a_i)}\,,\qquad H(x)\,\equiv\,\prod_{n=1}^\infty\left[
e^{-\frac{x^2}{n}}\,\left(1+\frac{x^2}{n^2}\right)^n\right]\,.
\ee
The specific form of the one-loop correction is directly related to the vanishing of the beta function for the gauge coupling. When the number of hypermultiplets differs from $2N$, a divergence ensues in the one loop fluctuation determinant above, which then renormalizes the coefficient in front of the quadratic term in the action i.e. the gauge coupling.  

We are only interested in evaluating Wilson loops at large-$N$ with $\lambda\equiv g^2N$ fixed. Therefore, what is needed is the large-$N$ eigenvalue distribution $\rho_\lambda(x)$ which solves the saddle point equation
\bea
&&\frac{8\pi^2}{\lambda}\,x- K(x)\,=\,{\cal P}\int_{-\mu}^\mu dy\,\rho_\lambda(y)\,\left(\frac{1}{x-y}-K(x-y)\right)\,,\quad x\in[-\mu,\mu]\label{spt}
\\\nonumber\\\nonumber
&&\int_{-\mu}^\mu \rho_\lambda(x)\,dx=1\,,\qquad K(x)\,\equiv\,
-\frac{H'(x)}{H(x)}\,.
\eea
Recall that the spectral density of the matrix model is defined as the large-$N$ `continuum' limit of
\be
\rho_\lambda(x)\,\equiv\,\frac{1}{N}\sum_{i=1}^N
\delta(x-a_i)\,.
\ee
An exact solution of the saddle point equation is not (yet) known, but crucial properties of $\rho_{\lambda}(x)$ can be inferred from the behaviour of function $K(x)$ which appears both as a central force term and in the pairwise interaction of eigenvalues. As observed in \cite{pz}, for small $x$, $K(x)\approx 2\zeta(3)x^3$ while for large $x$, $K(x)\to 2x \ln x $. This implies that  the pairwise interaction between eigenvalues is repulsive at short separation (dominated by Vandermonde repulsion) and attractive at very large eigenvalue separation. On the other hand  the central quadratic potential (attractive) dominates at short distances, but is overwhelmed by the repulsive $K(x)$ at large distances. Importantly, at large distances the one-body term $K(x)$ precisely counteracts the two-body force $K(x-y)$, so that for a large enough spread of the eigenvalue distribution, the behaviour at the endpoints is controlled by the Vandermonde repulsion and the quadratic one body potential.

We list below the main consequences \cite{pz} of these observations:
\begin{itemize}
\item {The spectral density associated to the eigenvalues of the matrix model \eqref{mm} above, attains a limiting form in the limit of infinite 't Hooft coupling, which is independent of the coupling,
\be
\rho_{\infty} (x)\,=\,\frac{1}{2}\,\frac{1}{\cosh\left(\frac{\pi x}{2}\right)}\,.
\label{limiting}
\ee
The spread of the eigenvalues is infinite ($\mu\to\infty$).
An important feature of the limiting form of $\rho_{\infty}(x)$ is that it cannot be used to yield a finite expectation value for the Wilson loop in the fundamental representation. In particular,
$\langle W_\Box\rangle\big|_{\lambda\to \infty}\,=\,
\int_{-\infty}^\infty dx\,\rho_{\infty}(x)\, e^{2\pi x}$ is divergent at face value.
}

\item{Hence, the finite $\lambda$ corrections to the exponential tail of the limiting distribution are crucial for determining the correct value of the Wilson loop at strong coupling. For finite (but large) $\lambda$, since the force on an eigenvalue at large distances is determined by the quadratic one-body potential and the Vandermonde repulsion, it can be argued that near its endpoints the eigenvalue distribution should smoothly interpolate between the limiting distribution and the Wigner semi-circle law,
\be
\rho_\lambda(x)\simeq\frac{8\pi}{\lambda}\sqrt{\mu^2-x^2}\,,
\qquad \qquad {x\sim\mu\gg1}.
\label{endpoint}
\ee
The location of the endpoint $\mu$ can be estimated by requiring the interpolating distribution to be correctly normalized i.e.,
$\int_{-\mu}^\mu dx\,\rho_\lambda(x)=1$, assuming that the crossover between \eqref{limiting} and \eqref{endpoint} occurs at $x\sim {\cal O}(1)$. This implies,
\be
\lambda \sim \sqrt\mu\, e^{\pi\mu/2}\,,\qquad\qquad \mu =\frac{2}{\pi}\ln\lambda+\ldots.
\ee
}
\item{ It is then straightforward to infer the $\lambda$-dependence of $\langle W_\Box\rangle$ in the large-$\lambda$ regime. The relevant integral is dominated by the endpoint of the spectral density, i.e., Eq.\eqref{endpoint}, so that
\be
\langle W_\Box \rangle\,=\, K\,\frac{\lambda^3}{{\ln\,\lambda}^{3/2}}\,.
\label{large}
\ee
Comparison with the standard result \footnote{For large-$N$ superconformal gauge theories with a supergravity dual (e.g. ${\cal N}=4$ SYM) , the Wilson loop $W \sim \exp(\sqrt\lambda A)/\lambda^{3/4}$. Here $A$ is the area of a minimal string (disk) world-sheet in $AdS$, whose boundary traces out the Wilson loop in the dual gauge theory. The pre-factor of $\lambda^{-3/4}$ arises from gauge-fixing on the world-sheet and depends on the number of zero modes, determined by the Euler character of the world-sheet \cite{sz}.  }
for the Wilson loop at strong 't Hooft coupling, in theories with  weakly curved AdS duals 
\cite{maldacena, malwil, reyyee,dgo,esz, dg}, then suggests that the effective string tension is $T_{\rm eff} = \tfrac{3}{2\pi}\ln\lambda$ in a putative string dual of the $N_f=2N$ theory.
}

\end{itemize}

 \section{High rank $k\sim {\cal O}(N)$ Wilson loops}

In the context of the AdS/CFT correspondence it has been known for some time  that Wilson loops in tensor representations of rank $k$ with $k \sim{\cal O}(N)$, in the large-$N$ limit, are computed by semiclassical D-brane probes in the (weakly curved) dual geometry \cite{drukkerfiol, yamaguchi1, polloops, yamaguchi2, Gomis:2006sb, mm} of the form $AdS_5\times X^5$. The antisymmetric and symmetric tensor representations are computed by $k$ fundamental strings `blown up' into probe D5- and D3-branes, respectively. The former are probes of the internal geometry and their dependence on the ratio $\frac{k}{N}$ is determined essentially by the volume of the four-cycle inside $X^5$, wrapped by the puffed up D5-brane. On the other hand, the D3-branes are embedded completely in the $AdS_5$ directions with world-volume $AdS_2\times S^2$.

It is conceivable that the behaviour of such high rank Wilson loops in the ${\cal N}=2$ superconformal theory will contain some hints of a large-$N$ string dual. It is {\it a priori} unclear whether such a string dual will have weak curvatures or not, but we expect that exact results from field theory may allow us to draw some inferences.

\subsection{Antisymmetric representation}
Given that supersymmetric Wilson loops in the ${\cal N}=2$ SCFT are computed by Pestun's matrix model \eqref{mm}, Wilson loops  transforming in various representations 
can be expressed as expectation values of appropriately symmetrized polynomials of eigenvalues of the random matrix $e^{2\pi a}$. In particular, 
the rank $k$ anti-symmetric tensor representation is explicitly
\be
\frac{\langle W_{A_k}\rangle_{{\cal N}=2}}{{\rm dim}(A_k)}\,=\,
\frac{1}{{\rm dim}(A_k)}\sum_{1\leq i_1 < i_2<\ldots< i_k\leq N}\,\langle\exp\left[2\pi(a_{i_1}+a_{i_2}+\ldots a_{i_k}) \right] \rangle_{\rm mm}\,,
\label{anti}
\ee
where the expectation value on the right hand side is within the random matrix model \eqref{mm} and the result is normalized by the dimension of the representation, ${\rm dim}(A_k) = \binom{N}{k}$. 

To evaluate this, we will follow the technique of \cite{mm}, assuming that the insertion of the Wilson loop operator does not change the large-$N$ eigenvalue distribution of the matrix model itself. This assumption is consistent in the strict large-$N$ limit, since the matrix model action is ${\cal O}(N^2)$, whilst the insertion of the rank $k$ Wilson loop operator introduces a `perturbation' of order $N$ and cannot alter the large-$N$ saddle point distribution. 

Since our primary interest is in the strong coupling limit of the ${\cal N}=2$ gauge theory,  we will compute the expectation value \eqref{anti} simply by using the limiting form of the eigenvalue density $\rho_{\infty}(x)$. Despite the fact that this distribution cannot provide a finite result for the fundamental representation, we will find that large enough antisymmetric tensor representations ($\frac{k}{N}$ fixed as $N\to\infty$) {\it can} converge to a smooth result (at infinite $\lambda$), independent of the 't Hooft coupling $\lambda$. (Independently of this, the Wilson loops for such large rank representations will always scale exponentially with $N$).

The Wilson loops in the antisymmetric representation  
can be read off as coefficients of the characteristic polynomial for the matrix $e^{2\pi a}$:
\be
\langle W_{A_k}\rangle\,=\,\oint \frac{dt}{2\pi i}\,\frac{1}{t^{N-k+1}}\,\langle\prod_{j=1}^N(t+e^{2\pi a_j})\rangle_{\rm mm}\,.
\ee
The large-$N$ limit provides a particularly convenient representation of this expression as an integral over the auxilliary spectral parameter $t$, which can then be evaluated in a saddle point approximation,
\be
\langle W_{A_k}\rangle\,=\,\oint \frac{dt}{2\pi i}\,{t^{k-1}}\,
\exp\left[N\,\int_{-\mu}^\mu dx\,\rho_{\lambda}(x)\,
\ln(1+e^{-2\pi x}\,t^{-1})\right]\,.
\label{integral1}
\ee

Note that this integral representation takes in the eigenvalue distribution (at infinite $N$) as input, and  can be used, in principle, to evaluate higher rank Wilson loops for any $\lambda$. 
It also enjoys a symmetry under the operation $k\to N-k$, which can be understood by performing the variable change $t\to t^{-1}$ and from the symmetry of $\rho_\lambda(x)$. This symmetry is 
 equivalent to charge conjugation, so that $W_{A_k}=W_{A_{N-k}}$ as expected for sources transforming in the antisymmetric tensor representation of $SU(N)$. 

\begin{figure}[h]
\begin{center}
\epsfig{file=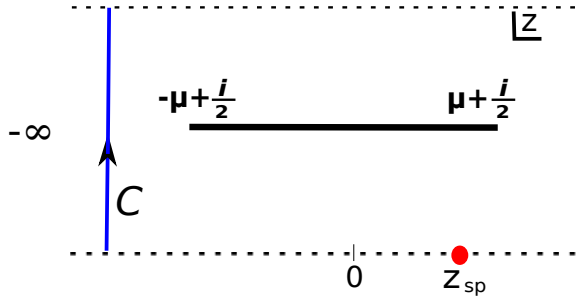, width =3in}
\end{center}
\caption{\small{ The integral along the contour $C$, on the cylinder yields the rank  $k$ antisymmetric Wilson loop. The contour lies to the left of the branch cut between 
$-\mu +\frac{i}{2}$ and $\mu +\frac{i}{2}$ where $\mu\simeq\tfrac{2}{\pi} \ln\lambda$. The saddle point on the first sheet lies on the real axis at $z_{\rm sp}$, well away from the branch cut.
}}
\label{integral1fig} 
 \end{figure}

Another feature of the formula \eqref{integral1}, is that the exponent (as a function of $t$) has a branch cut singularity. This branch cut arises from the $N$ zeroes of the characteristic polynomial coalescing in the large-$N$ limit to form a continuum. The branch points are at $t= - e^{-2\pi\mu} \sim - \lambda^{-4}(\ln\lambda)^2$ and $t = -e^{2\pi\mu} \sim \lambda^4(\ln\lambda)^{-2}$. In the limit of infinitely large 't Hooft coupling the branch cut stretches along the entire negative $t$-axis, and we must be careful whilst  considering this limit to evaluate  the Wilson loop.

In the limit $k, N\to\infty$ with $\frac{k}{N}$ fixed, the $t$-integral  can be evaluated by the method of steepest descent. The condition for the existence of a saddle-point in the $t$-plane is then,
\be
\int_{-\mu}^\mu dx\,\frac{\rho_{\lambda}(x)}{1+e^{ 2\pi x}t}\,-f\,=\,0\,,\qquad f\,\equiv\,\frac{k}{N}\,.
\label{saddleeq}
\ee
An interesting and potentially useful interpretation of the system can be obtained by adopting a different parametrization,
\be
t\,\equiv\, e^{- 2\pi z}\,.
\ee
This maps the plane to the cylinder with a branch cut along $-\mu+\frac{i}{2}\leq z\leq +\mu+\frac{i}{2}$ (Figure \eqref{integral1fig}). Along the branch cut the integrand develops an imaginary part. Given the analytic structure of the integrand, we can use this procedure as long as the location of the saddle points is away from the branch cut singularity. With the exponential parametrization, the saddle point equation acquires a nice interpretation,
\be
\int_{-\mu}^\mu dx\,\frac{\rho_{\lambda}(x)}{1+e^{2\pi (x-z)}}\,=\,f\,.
\label{fermion}
\ee
This equation gives the occupation number of fermions in the presence of a chemical potential $z$ and with a density of states $\rho_\lambda$. Depending on the specific form of $\rho_\lambda(x)$, we may be able to infer an effective 
``temperature'' of the system by appropriate rescalings of the integration variable.

\subsection{Infinite $\lambda$ limit}
The  large-$\lambda$ distribution function has the property that its endpoints run off to infinity in the limit of infinite 't Hooft coupling. We know that the fundamental Wilson loop  
\eqref{large} diverges in this limit because its expectation value is dominated by the largest eigenvalues in the vicinity of the endpoint of the eigenvalue density $\rho_\lambda$.
However, we can expect large rank representations to behave  differently. While individual terms in the sum \eqref{anti} may diverge as $\lambda$ is taken to infinity, in the limit of large $k, N$ with $\frac{k}{N}$ fixed, the total number of terms in the sum 
grows exponentially with $N$. As the distribution $\rho_\infty$ is peaked about $x=0$, a majority of eigenvalues reside in the bulk of the distribution and their contribution to the sum can dominate the result for high rank representations. Indeed, this is precisely what occurs. 

The integral representation for the antisymmetric loop at $\lambda =\infty$ explicitly reads,
\be
\langle W_{A_k}\rangle\big|_{\lambda\to\infty}\,=\,i\oint_{C} 
dz\,
\exp\left[N\left(2\ln\left(\cosh\tfrac{\pi z}{2}+\tfrac{1}{\sqrt 2}\right)-\left(f-\tfrac{1}{2}\right)\,2\pi z+\ln 4\right)\right]\,.
\label{integral2}
\ee
Despite the fact that the integrand in the original expression 
\eqref{integral1} for $W_{A_k}$ is manifestly periodic under $z\to z+ i$, its explicit form above in the $\lambda\to\infty$ limit, is only periodic under the shift $z\to z+ 4 i$. This is due to the branch cut shown in Figure\eqref{integral1fig} which now has an infinite extent. 

The large-$N$ saddle point equation follows directly from the above result,
\be
%2-\sqrt 2\,t^{1/4}+\sqrt 2\, t^{3/4}-2\,(1+t)\,f\,=\,0\,.
 \frac{1}{2}\sinh\tfrac{\pi z}{2}\,\left(\cosh\tfrac{\pi z}{2}+\tfrac{1}{\sqrt 2}\right)^{-1}\,= \,f-\frac{1}{2}\,.
\label{eq1}
\ee
This equation has distinct roots with ${\rm Im}\,z=0$ and 
${\rm Im}\,z = 2 i$. For the moment, we focus our attention only on the real root. The complex root lies on a different sheet, due to the branch cut discussed above. We find one saddle point on the real axis satisfying,
\be
\coth\left(\tfrac{\pi}{2}\,z_{\rm sp}\right)\,=\,
\frac{1-\sqrt{\tfrac{1}{2}- \left(f-\tfrac{1}{2}\right)^2}}{f-\tfrac{1}{2}}.
\ee
As a function of $f$, in the domain $0\leq f\leq 1$, $z_{\rm sp}$ takes on all values on the real axis. For small $f$ and $f$ approaching unity, the saddle point runs of to infinity,
\bea
z_{\rm sp}&\,=\,&\tfrac{2}{\pi}\ln f+\ldots\,,\qquad \qquad\quad f \ll 1\,,
\\\nonumber
&\,=\,&-\tfrac{2}{\pi}\ln(1-f)+\ldots\,,\qquad(1-f) \ll 1\,.
\eea

A nice property of this solution as a function of $f$, is that
\be
z_{\rm sp}(f)\,=\, - z_{\rm sp}(1-f)\,.
\ee
This reflects the symmetry of the integral representation for the antisymmetric Wilson loop, under $z\to - z$ accompanied by the replacement $f\to 1-f$. 
In the free fermion interpretation, $z_{\rm sp}$ acts as a real chemical potential. Since the distribution $\rho_\infty(x)$ is independent of $\lambda$, the effective temperature for the Fermi-Dirac distribution in Eq.\eqref{fermion} is ${\cal O}(1)$.

We now proceed to evaluate the Wilson loop itself. After deforming the contour to pick up the saddle point contribution 
(formally, before taking the $\lambda\to\infty$ limit) from the first sheet, we have
\be
\langle W_{A_k}\rangle\big|_{\lambda\to\infty}\,\approx\,
%\frac{1}{\rm dim(A_k)}
%\oint \frac{dt}{2\pi i}\,
\exp\left[N\left(2\ln\left[\frac{\sinh\left(\tfrac{\pi }{2}z_{\rm sp}\right)}{f-\tfrac12}\right]-(f-\tfrac12)\,2\pi z_{\rm sp}\right)\right]\,.
\label{nice}
\ee
%Picking up the saddle point contribution at $t={\cal T}_+$, we find
%\bea
%%\exp\left[N\,\left(2\ln\left(\frac{1 + 2 f + \sqrt{1 + 4f (1 - f)}}{4\,f^2}\right)\right.\right.\\\nonumber
%&&\left.\left.-4(1-f)\ln\left(\frac{1 - 2 f + \sqrt{1 + 4f (1 - f)}}%{2\sqrt 2\,f}\right)\right)\right]\right|_{t={\rm cal T}(f)}\,
%\eea
The explicit expression is not particularly illuminating, but the expansion of the exponent in a power series in $f =\tfrac {k}{N}$ is interesting:
\be
\boxed{
\ln\,\langle W_{A_k}\rangle\big|_{\lambda\to\infty}
\,=\, N\left[- 4\,\frac{k}{N}\,\ln\left(\frac{\sqrt2}{e} \frac{k}{N}\right)-\frac{8}{3}\left(\frac{k}{N}\right)^3 + 4 \left(
\frac{k}{N}\right)^4\ldots\right]\,}\qquad \frac{k}{N}\ll 1\,.
\label{smallf1}
\ee

The most notable features of this 
formula\footnote{Note that our definition of $W_{A_k}$, Eq. \eqref{anti}, does not include the normalization factor given by the dimension of the representation.} are the non-analytic (logarithmic) dependence on $f$ for small $f$, the complete absence of any dependence on the 't Hooft coupling $\lambda$ and that beyond the leading order, subsequent terms are organized in a power series in $\frac{1}{N}$ (as opposed to $\frac{1}{N^2}$, for instance). The logarithmic behaviour  is an immediate consequence of Eq.\eqref{nice} and that 
$z_{\rm sp} \sim  \ln f$ for small $f$.

\subsection{Large but finite $\lambda$}
For any finite value of the coupling constant, the eigenvalue distribution has a finite extent \cite{pz}, and the exponential tail of $\rho_\lambda(x)$ smoothly matches on to a Wigner semi-circle  distribution which vanishes at $|x|= \mu\approx\tfrac{2}{\pi}\ln \lambda+\ldots$. In the fixed $k$, large-$N$ limit, we expect that the value for the (exponent of) the antisymmetric Wilson loop is $k$ times that of the fundamental loop (from large-$N$ factorization). The result for the fundamental loop \eqref{large} depends on $\lambda$, whereas we have seen that for large enough representations with $f=\frac{k}{N}$ fixed, the observables have a $\lambda$-independent limit at large $\lambda$. We will now see how these two results can be reconciled.

\begin{figure}[h]
\begin{center}
\epsfig{file=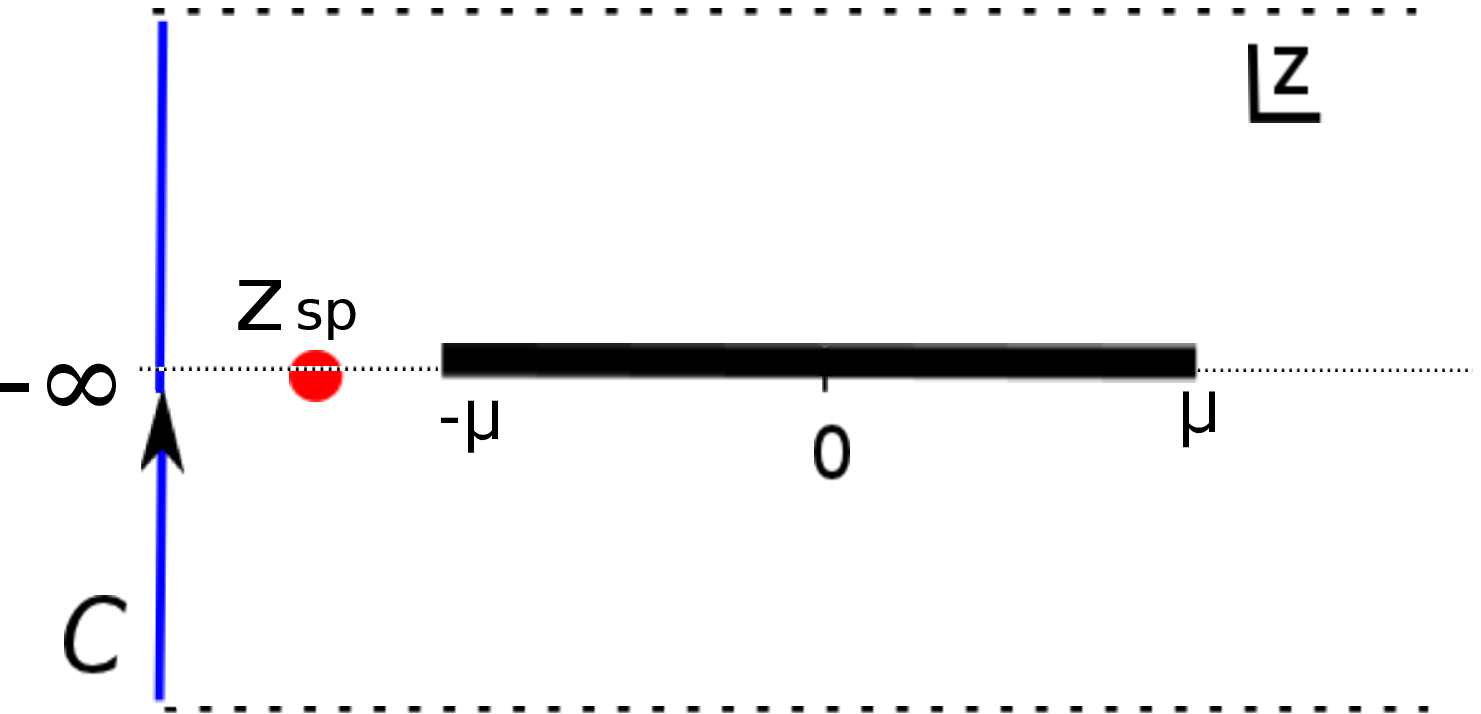, width =2.3in}
\hspace{0.7in}
\epsfig{file=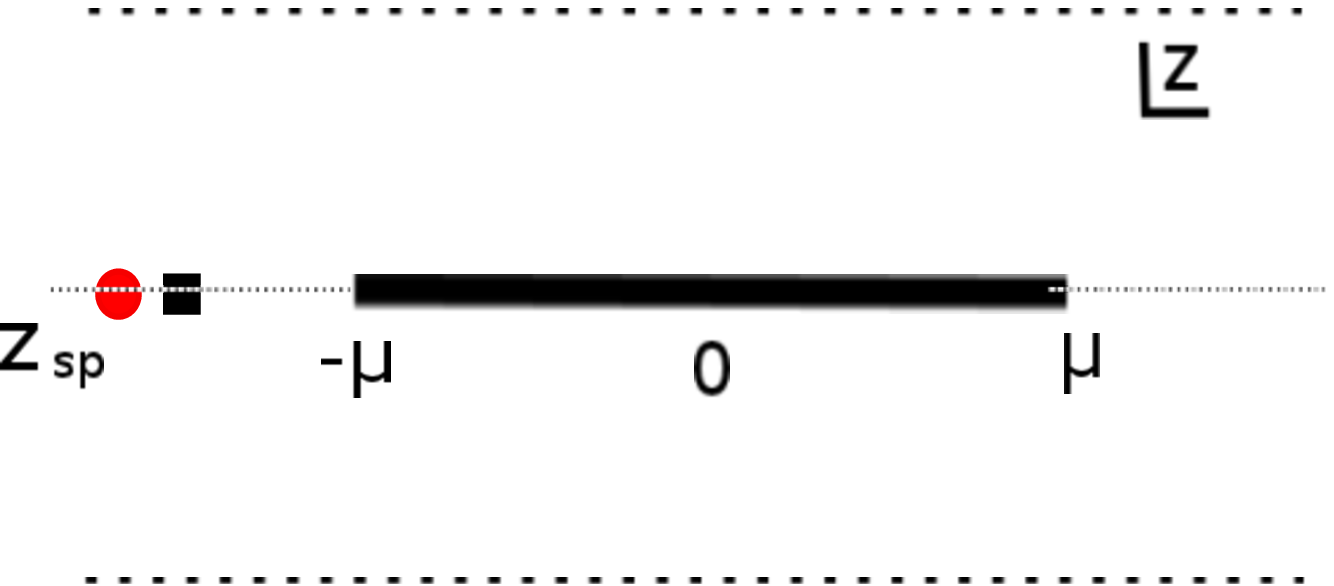, width =2.4in}
\end{center}
\caption{\small{ {\underline {\bf Left}}: The branch cut singularity associated to the integral representation of the symmetric Wilson loop lies on the real axis in the $z$-plane. For $f$ less than a critical value $f_c$ the saddle point $z_{\rm sp}$ also lies on the real axis. {\underline{\bf Right}}: When $f>f_c$ the saddle crosses the branch point. Beyond, a qualitatively new saddle point dominates the integral: a single eigenvalue 
well-separated from the continuum, reminiscent of Bose condensation. A similar configuration in ${\cal N}=4$ SYM was related by \cite{yamaguchi1} to the position of a probe  D3-brane in $AdS_5\times S^5$.
}}
\label{integral2fig} 
 \end{figure}

To understand that there must be a qualitative change in the behaviour, we do not need the explicit form of $\rho_\lambda$. All we require is that the eigenvalue distribution has a finite extent for finite 't Hooft coupling. In particular we know that the distribution is non-zero only in the domain $-\mu < x < \mu$. The relevant question is whether the solution to Eq.\eqref{saddleeq} is sensitive to this finite extent of the distribution. When $f$ is taken to be sufficiently small, we can find a saddle point with $e^{-2\pi z}\gg e^{2\pi \mu}$, by expanding \eqref{saddleeq} in powers of $e^{2\pi z}$ and keeping the leading term
\be
e^{-2\pi z} \simeq \frac{1}{f}\,\int_{-\mu}^\mu dx\,\rho_\lambda(x)\,e^{-2\pi x}
=\frac{1}{f}\,\langle W_\Box\rangle\,,\qquad e^{-2\pi z}\gg e^{2\pi\mu}\,.
\ee
We already know that $\langle W_\Box\rangle\sim e^{3\pi\mu\,/2}\sim
\lambda^{3}/(\ln\lambda)^{3/2}$ at strong coupling, and therefore the above solution applies for small $f$ such that
\be
f\,\ll \,e^{-\pi\mu/2}\,\sim\, \lambda^{-1}\,\sqrt{\ln\lambda}\,\ll \,1\,. 
\ee
Substituting the small $f$ saddle point into the integral representation for the antisymmetric Wilson loop, we obtain
\be
\frac{\langle W_{A_k}\rangle}{{\rm dim}(A_k)}\quad \to\quad
\frac{k!(N-k)!}{N!}\,\exp\left(N\left[f+ f\ln\left(f^{-1}\langle W_\Box\rangle\right)\right]\right)\,=\,\langle W_\Box\rangle^k\,.
\label{asymsmallf}
\ee
Note that the overall normalization given by the dimension of the representation is precisely cancelled by  contributions to the exponent at the saddle point. Therefore, explicitly, we have
\be
\langle W_{A_k}\rangle \sim \frac{\lambda^{3k}}{(\ln\lambda)^{3k/2}}\,, \qquad {\rm for}\quad f\ll e^{\pi\mu/2}\,.
\label{smallf2}
\ee 
Hence, the antisymmetric Wilson loop at strong coupling exhibits a crossover from $\lambda$-dependent behaviour \eqref{smallf2} 
for $\tfrac{k}{N}\ll\lambda^{-1}\,\sqrt{\ln\lambda} $ to a $\lambda$-{\em independent} limit \eqref{smallf1} for parametricaly larger $f=\frac{k}{N}$\,. We do not know if the cross-over is actually smooth, but given that the  saddle-point $z_{\rm sp}$ (see figure\eqref{integral2fig}) is located well away from the branch points, we do not expect non-analyticities in the associated configuration. Furthermore, we have not seen evidence of more than one saddle point, so a first order transition due to competition between two or more configurations appears unlikely. Settling the  actual nature of the crossover behaviour described above will require a more detailed analytical or numerical understanding of the finite-$\lambda$ eigenvalue distribution.

\section{Comparison with ${\cal N}=4$ SYM}
A comparison of the result above with the corresponding one for ${\cal N}=4$ SYM is quite useful. Circular Wilson loops in  ${\cal N}=4$ SYM are computed by the Gaussian matrix model with eigenvalue density
\be
\rho_\lambda(x)\,=\ \frac{2}{\lambda}\sqrt{\lambda-(2\pi x)^2}\,.
\ee
Using this eigenvalue distribution and upon rescaling variables in Eq.\eqref{fermion}, $\tilde x\equiv \tfrac{\sqrt{\lambda}}{2\pi}\,x$ and $\tilde z\equiv\tfrac{\sqrt{\lambda}}{2\pi}\,z$, the saddle point value of the rank $k$ antisymmetric Wilson loop is determined by the condition, 
\be
\frac{2}{\pi}\int_{-1}^1 d\tilde x\,\frac{\sqrt{1-\tilde x^2}}{1+e^{\sqrt{\lambda}(\tilde x-\tilde z)}}\,=\,f\,.
\label{n=4anti}
\ee
The crucial difference with respect to the ${\cal N}=2$ theory studied previously, is the explicit dependence on the 't Hooft coupling of the ${\cal N}=4$ theory, which now plays the role of the temperature of the Fermi-Dirac distribution.

The limit of infinitely strong coupling ($\lambda\to\infty$) can be interpreted as a ``zero temperature'' limit for free fermions and therefore all Fermi levels $\tilde x$ below the chemical potential $\tilde z$ are ``occupied'', whilst all $\tilde x>\tilde z$ remain ``empty''. Hence, the saddle point $\tilde z_{\rm sp}$ is determined by setting the total number of fermions (eigenvalues) in the ground state to $k$
\be
\frac{2}{\pi}\int_{-1}^{\tilde z_{\rm sp}}dy\,\sqrt{1-\tilde x^2}\,=\,\frac{k}{N}\,.
\ee
%Interestingly, this strong coupling saddle point actually lies on the eigenvalue distribution. This is in contrast to the general 
%situation for the ${\cal N}=2$ SCFT (see Figure\eqref{integral1fig}).
This leads to the known result \cite{yamaguchi1, yamaguchi2, mm},
\be
\langle W_{A_k}\rangle_{{\cal N}=4}\big|_{\lambda \to\infty}\,=\,
\exp\left(N\,\frac{2\sqrt\lambda}{3\pi}\sin^3\theta_k\right)\,,\qquad
\frac{1}{\pi}(\theta_k -\sin\theta_k\cos\theta_k)\,=\,\frac{k}{N}\,.
\label{strongn=4}
\ee
This result is only valid when $\lambda\to\infty$. 
For any finite $\lambda$, our free fermion interpretation above shows that there will be corrections due to occupation of all levels above the chemical potential. 
It would be interesting to look at the detailed form of the corrections to the infinite coupling limit. The leading corrections include powers of $\lambda^{-1}$ and exponentially suppressed terms $\sim e^{-\sqrt\lambda(1-|\tilde z_{\rm sp}|)}$ which are suggestive of world-sheet instanton corrections to the D5-brane saddle point. In the free Fermi picture, these corrections can be thought of as contributions from excited levels at a finite temperature of order $\lambda^{-{1/2}}$. 

The salient features of the formula Eq.\eqref{strongn=4} are the dependences on $N$ and $\sqrt\lambda$. The exponential growth with $N$ is due to the large $k$ limit with $k\sim {\cal O}(N)$, while the factor of $\sqrt
 \lambda$ is a consequence of the large 't Hooft coupling. In the IIB string dual description on $AdS_5\times S^5$, the dependence on $N$ and $\sqrt\lambda$  arises straightforwardly from the tension of a semiclassical D5-brane wrapping a flux supported $S^4\subset S^5$, which computes the antisymmetric Wilson loop \cite{yamaguchi1,yamaguchi2, polloops, Gomis:2006sb}. The D5-brane tension $T_{\rm D5}=1/(2\pi g_s\alpha^{\prime3})$ expressed as a function of the 't Hooft coupling and AdS radius $R_{\rm AdS}$ becomes ${N\sqrt\lambda/8\pi^4 R^6_{\rm AdS}}$. Since all the curvature scales are set by $R_{\rm AdS}$, the scaling of the Wilson loop is completely determined.
 The angular variable $\theta_k$ introduced above determines a specific latitude in $S^5$ and the supergravity result precisely matches  Eq.\eqref{strongn=4}. 

The result for the ${\cal N}=2$ SCFT  suggests that the antisymmetric Wilson loop for large $k$, in a putative large-$N$ string dual, should indeed computed by a D-brane since its action is ${\cal O}(N) \sim g_s^{-1}$, where $g_s$ is the string coupling. However, the absence of any dependence on the 't Hooft parameter $\lambda$ also indicates that this D-brane must probe (internal) portions of the geometry with string scale curvatures. For sufficiently small $\frac{k}{N}$ the antisymmetric loop is $k$ times the fundamental Wilson loop, which, due to its growth with $\lambda$ is likely computed by a string in $AdS_5$ with a large  effective tension $T_{\rm eff}=\tfrac{3}{2\pi}\ln\lambda$. The transition/crossover from this semiclassical picture to a less familiar situation could be driven by $k \sim {\cal O}(N)$ strings ``puffing up'' into a D-brane probing highly curved or non-geometric portions of the dual string theory.

\section{Symmetric Representation}
The symmetric tensor representations can be obtained by considering a generating function which is the inverse of the characteristic polynomial,
\bea
\langle W_{S_k}\rangle\,&=&\,\oint \frac{dt}{2\pi i}\,t^{k-1}\,\left\langle\frac{1}{\prod_{i=1}^N(1- t^{-1}\,e^{2\pi a_i})}\right\rangle_{\rm mm}\,,\label{integral2}\\\nonumber
&\to&
i\oint_C dz\,
\exp\left[-N\,\int_{-\mu}^\mu dx\,\rho_{\lambda}(x)\,
\ln(1 - e^{-2\pi (x-z)})-2\pi k z\right]\,.
\eea
These formulae pick out a symmetrized polynomial in the eigenvalues $e^{2\pi a_i}$,
\be
\frac{\langle W_{S_k}\rangle_{{\cal N}=2}}{{\rm dim}(S_k)}\,=\,
\frac{k! (N-1)!}{(N+k-1)!}\sum_{1\leq i_1 \leq i_2<\ldots\leq i_k\leq N}\,\langle\exp\left[2\pi(a_{i_1}+a_{i_2}+\ldots a_{i_k}) \right] \rangle_{\rm mm}\,.
\label{sympol}
\ee
The main difference between this and the antisymmetric tensor representation is that now the eigenvalues $a_i$ appearing in each term of the polynomial, need not all be distinct. In fact, we may separate out the terms in two categories: those with $i_1\neq i_2\neq\ldots\neq i_k$, and those where some of the $i_\ell$ coincide. The contribution from the former is identical to the antisymmetric representation; the latter includes terms where most eigenvalues coincide so that their behaviour is like a multiply wound loop. In fact we will see below that the integral representation is dominated by two distinct saddle points, related precisely to the two categories of terms in the symmetrized polynomial. At strong coupling the saddle point related to the multiply wound loop grows exponentially with the 't Hooft coupling and dominates the symmetric loop.

The integral representation above which leads to the large $N$ limit, involves a function with a branch cut along the real axis $-\mu\leq z\leq \mu$. Therefore, we need to first determine whether putative saddle points that contribute to the integral, lie in the vicinity of the branch cut on the real axis.
Now the large-$N$ saddle point equation can be interpreted as
fixing the number of bosons at a chemical potential $z$, with a density of states given by $\rho_\lambda(x)$, 
\be
\int_{-\mu}^\mu dx\,\frac{\rho_{\lambda}(x)}{e^{2\pi (x-z)}-1}\,
=\,f\,.
\label{saddleeq1}
\ee
Formally, the equation for the symmetric representation 
can be obtained from Eq.\eqref{saddleeq} after the replacement 
$f\to -f$ and $z\to z+\tfrac{i}{2}$.  However, now the analogy with the Bose-Einstein distribution implies that the system could display a phase transition analogous to Bose-Einstein condensation as the chemical potential is dialled from low to high values.

\subsection{Large $\lambda$ and a non-analyticity}
 
\paragraph{\underline {Small $f$ solution}:}
We begin by locating the solution to Eq.\eqref{saddleeq1}, with the finite $\lambda$ distribution function.
It is clear that the equation is likely to have solutions  with  $z < - \mu$, for then the integrand is positive definite. A quick check will now confirm this expectation.
Assuming that the equation is solved by some $e^{2\pi z}\ll e^{-2\pi\mu}$, and  expanding Eq.\eqref{saddleeq1} to the lowest order in $e^{2\pi z}$, we find a saddle point given by
\be
\exp(-2\pi z_{\rm sp})\approx \frac{2 e^{3\pi\mu/2}}{3\pi f}\,\implies 
\langle W_{S_k} \rangle \propto \frac{\lambda^{3k}}{(\ln\lambda)^{3/2}}\,,\qquad f\ll\lambda^{-1}\sqrt{\ln\lambda}\ll 1\,.
\label{smallsymm}
\ee
This is identical to the antisymmetric loop in the same limit, and is consistent with the view that in the fixed $k$ large $N$ limit, the two should be equivalent to $k$ coincident Wilson loops in the fundamental representation. In fact, we can be more precise. Following the same arguments that led to Eq.\eqref{asymsmallf}, we find
\be
\frac{\langle W_{S_k}\rangle}{{\rm dim}(S_k)}\quad\to\quad
\langle W_\Box\rangle^k\qquad{\rm for}\quad f\ll\lambda^{-1}\sqrt{\ln\lambda}\ll 1\,.
\ee
What is now interesting in the case of the symmetric representation is that for a given (large) $\lambda$, as $f$ is increased smoothly, the saddle point  moves to larger values of $z$, towards the branch-point at $z=-\mu$. Beyond this, the nature of the saddle point must change as it moves through the branch cut joining
 $z=-\mu$ and $z= +\mu$. Let us now ascertain the critical value  $f=f_c$ at which this occurs. A useful way to proceed  is to first formally expand Eq.\eqref{saddleeq} as a power series in $e^{2\pi z}$, assuming $ z < -\mu$:
 \be
 \sum_{n=1}^\infty\, e^{2 n\pi z }\,
 \int_{-\mu}^\mu dx\,\rho_{\lambda}(x)\, e^{- 2\pi n x }
 \,=\,f\,.
 \ee
 The coefficients in this expansion are the VEVs of multiply wound loops in the fundamental representation. Following the logic of the arguments presented in \cite{pz}, these should be determined completely by the endpoint of the eigenvalue distribution at strong coupling i.e. the endpoint of the Wigner semi-circle law \eqref{endpoint}. At strong coupling ($\mu\gg 1$), we have
 \be
 \int_{-\mu}^\mu\rho_\lambda(x)\, e^{-2\pi n x}\,\simeq\,R\, \frac{\sqrt{\mu}}{\lambda}\,\,\frac{e^{2n\pi\mu}}{n^{3/2}}\,,
 \ee
 where the right hand side follows completely from the square root behaviour near the endpoints of the eigenvalue distribution. This assumption yields $R=2$; a more careful estimate of $R$ performed in \cite{pz}, gave $R=2.18$ (for $n=1$). We note that in principle $R$ could depend on $n$, but we take this dependence to be weak. With increasing $n$, the integral is more sharply peaked near the endpoint at $x=\mu$ and we expect that taking $R=2$ becomes a better approximation. The strong coupling saddle point equation can then be rewritten (assuming $e^{2\pi z} <  e^{-2\pi\mu}$),
\be
R\,\tfrac{\sqrt\mu}{\lambda}
\,{\rm Li}_{\frac{3}{2}}\,(\,e^{2\pi(\mu+z)}\,) \,=\,f\,.
\ee
The critical value of $f$ for which the saddle hits the branch point at $z= -\mu$ is,
\be
f_c\,\simeq\, R\,\sqrt{\tfrac{2}{\pi}}\,\zeta\left(\tfrac{3}{2}\right)
\,\lambda^{-1}\,\sqrt{\ln\lambda}
\,.
\ee  
Pushing further the analogy with the boson gas, this phenomenon is reminiscent of the onset of Bose-Einstein condensation, as the chemical potential $z$ is dialled towards the lowest energy level (eigenvalue) at $x=-\mu$. At this point one should treat the lowest level (eigenvalue) and its occupation number separately, so that it is `split' from the higher levels. Motivated by this line of thinking we find a result quite similar to that of \cite{yamaguchi1} for the multiply wound Wilson loop in ${\cal N}=4$ SYM.
\\\\
\underline{\bf Large $f$ solution $(f>f_c)$:} 
Following the intuition gained from the free boson gas picture, we allow for the possibility that for $f> f_c$, the occupation number associated to the lowest eigenvalue $a_1$, at the edge of the distribution, is macroscopic
\be
n_1\,\equiv\,\frac{1}{e^{2\pi(a_1-z)}-1}\,\sim\,{\cal O}(N)\,.
\ee
This assumption alters the saddle point equations obtained from the matrix model \eqref{mm} for this eigenvalue alone, while leaving unaltered the large-$N$ distribution of the remaining $N-1$ eigenvalues.

To extract the behaviour of the symmetric Wilson loop for $f> f_c$, we turn to the discrete (finite $N$) version of the saddle point of the matrix integral \eqref{mm}, but in the presence of an insertion of the generating function for symmetric loops, i.e. Eq. \eqref{integral2}. In the large-$N$ limit (assuming $n_1\sim {\cal O}(N)$), we arrive at two saddle point equations upon varying with respect to $a_1$ and with respect to $t$:
\bea
&&\frac{8\pi^2}{\lambda}\,a_1 - K(a_1)-
\int_{-\mu}^\mu dx \,\rho_\lambda(x)\left(\frac{1}{a_1-x}-K(a_1 -y)\right)\,=\,- \pi \frac{n_1}{N}\,,\label{a}\\\nonumber\\
&& \frac{n_1}{N} + \int_{-\mu}^\mu dx\frac{\rho_\lambda(x)}{e^{2\pi(x-z)}-1} = f\,.
\label{b}
\eea
The rest of the $(N-1)$ eigenvalues continue to be governed by the unperturbed distribution $\rho_\lambda(x)$. To establish the exact form of the split eigenvalue solution for $f\gtrsim f_c$, just above the transition, we would need the explicit form of $\rho_\lambda(x)$ for finite, large $\lambda$. However, the exact solution of the large-$N$ equation \eqref{spt} is unknown at  finite $\lambda$. Despite this, it is easy to infer the  nature of this saddle point for large enough $\lambda$ or $\frac {k}{N}$, as we now see.

We look for a solution to the above set of equations by assuming that $z\ll -\mu$. We may then safely ignore the exponentially small contribution from the continuum in \eqref{b}, so that
\be
\frac{n_1}{N}\approx f\,\implies a_1\approx z+\frac{2\pi}{k}\ll -\mu\,. 
\ee 
Using the asymptotic form of $K(x)$ we find that the force on an eigenvalue, $a_1$, at large distances from the continuum distribution, is only due to the harmonic central potential. This must be balanced by the constant force on the right hand side of Eq.\eqref{a}, yielding,
\be
a_1\approx -\frac{\lambda} {8\pi}\frac{k}{N}\,.
\ee
This result is valid as long as,
\be
a_1\ll -\mu, \qquad{\rm or}\qquad \frac{\lambda}{8\pi}\frac{k}{N}\gg \frac{2}{\pi}\ln\lambda\,,
\ee
which is easily satisfied at strong coupling if we take
$f=\frac{k}{N}\sim{\cal O}(1)$. Notice that this implies a lower bound on $f$ for fixed large $\lambda$, which is safely above $f_c$ and so the solution above describes the Wilson loop in the correct `phase'.

Substituting our solution into Eq.\eqref{integral2}, and taking care to include the contribution from the quadratic potential term in Eq.\eqref{mm} we find
\be
W_{S}^{(1)}\,\approx\, \exp\left[{2 N \,\left(\sqrt\lambda \frac{k}{4 N}\right)^2}\right]\,.
\label{symresult}
\ee 
This result has corrections to its exponent, scaling as $\sim e^{-\frac{\lambda k}{4 N}}$. We have introduced a superscript to identify the contribution to the Wilson loop from this saddle point. Below we will encounter another strong coupling saddle point.

Interestingly, the entire analysis above could have been adapted to calculate the symmetric Wilson loop in ${\cal N}=4$ SYM, yielding exactly the same result as \eqref{symresult}, provided $\frac{k}{N}$ is taken to be large enough. Indeed, the known exact formula for that case (both from the probe D3-brane calculation \cite{drukkerfiol} and the Gaussian matrix model \cite{yamaguchi1,hep-th/0604209,mm}) reduces to \eqref{symresult} when $\sqrt{\lambda}\frac{k}{N}\gg 1$. These observations indicate that symmetric Wilson loops may be computed by a semiclassical object (D-brane) in the large-$N$ string dual of the ${\cal N}=2$ SCFT and that, in particular, the $AdS_5$ portion of the geometry probed by it may not be highly curved.

\subsection{A second strong coupling saddle point}
 
We found above that the saddle point contribution for the circular Wilson loop in the symmetric tensor representation grows exponentially with $\lambda$ (and with $N$) at large $\lambda$. This is in stark contrast to the antisymmetric loop. However, it turns out that there is another saddle point which contributes to the expectation value of the symmetric Wilson loop and is $\lambda$-independent at infinite $\lambda$. The quickest way to see this is to note that the integrand in \eqref{integral2} and its associated saddle point equation can be obtained, up to an overall sign, from those for the antisymmetric representation after the replacements: $z\to z+\frac{i}{2}$ and $f\to -f$. At this new saddle point
\bea
&& W_{S}^{(2)}\big|_{\lambda\to\infty}\,=\\\nonumber
&&\exp\left[- N\left(2\ln\left(\cosh\tfrac{\pi}{2} (z_{\rm sp}+\tfrac{i}{2})+\tfrac{1}{\sqrt 2}\right)+\left(f+\tfrac{1}{2}\right)\,2\pi (z_{\rm sp}+\tfrac{i}{2})+\ln 4\right)\right]\,,
\label{integral5}\\\nonumber\\\nonumber
&& \coth\left(\tfrac{\pi}{2}(z_{\rm sp}+\tfrac{i}{2})\right)\,=\,
\frac{\sqrt{\tfrac{1}{2}-\left(f+\tfrac{1}{2}\right)^2}-1}{f+\tfrac{1}{2}}\,.
\eea
Expanding the contribution for small $f =\frac{k}{N}$,
\be
W_{S}^{(2)}\big|_{\lambda\to\infty}\,=\,\exp N\left[-4 \frac{k}{N} \ln\left(\frac{\sqrt 2}{e}\frac{k}{N}\right)-\frac{40}{3}\left(\frac{k}{N}\right)^3-36\left(\frac{k}{N}\right)^4+\ldots\right]\,.
\ee
The full symmetric Wilson loop at large $N$, is given by the sum of the contributions from the two saddle points,
\be
\langle W_{S_k}\rangle\,=\,W_S^{(1)}+W_{S}^{(2)}\,.
\ee
Hence the symmetric Wilson loops can exhibit yet another type of non-analyticity, namely, a first order phase transition at large $N$ due to a competition between these two configurations. However, for fixed $f$ and large $\lambda$, it appears that $W_S^{(1)}$ will always dominate over the second saddle point, which remains exponentially suppressed.

The existence of two saddle points for the symmetric representation can be intuitively understood from  the symmetrized polynomial representation Eq.\eqref{sympol}. One configuration is dominated by the terms which compute the multiply wound Wilson loop (${\rm Tr} e^{2\pi k a }$) and these have $i_1=i_2=\ldots i_k$, whilst the second type of configuration is dominated by the terms with all or most eigenvalues being distinct and is therefore similar to the antisymmetric loop. A similar transition has also been noted in ${\cal N}=4$ SYM \cite{mm, hep-th/0604209}.

\section{Discussion}

The matrix model of \cite{pestun} provides  a powerful tool for extracting predictions for a large class of observables in ${\cal N}=2$ SCFTs. It may also serve as a potential window into large-$N$ string duals of SCFTs at strong coupling. The analysis in this note presents us with one class of observables in the $N_f=2N$ theory that could shed light on the corresponding large-$N$ dual. Below we discuss some related questions and directions for future study.

 \paragraph{\underline{Relation to Kondo models}}: Wilson loops in different representations can be viewed as impurity spins \cite{Gomis:2006sb} coupled to an ambient theory which, in our case, is an ${\cal N}=2$ SCFT.
These  may be regarded as supersymmetric versions of the Kondo model 
(see e.g. \cite{parcollet, Sachdev:2010uz}). An antisymmetric Wilson loop computes the action of a fermionic impurity interacting with the SCFT.
There is at least one SCFT namely, ${\cal N}=4$ SYM, for which the circular Wilson loop in the ambient theory at zero temperature, has been related to the impurity model (or Polyakov loop) at finite temperature \footnote{The origin of this relation has been explained  in \cite{muck} at strong coupling. In an SCFT the size of a circular loop, or the temperature, can always be rescaled to unity. It is not {\it a priori} clear that the circular loop at zero temperature should be related to the Polyakov loop at finite temperature since one is supersymmetric while the other has anti-periodic boundary conditions for ambient fermions around the thermal circle.} 
 \cite{muck, hkt}. 
 
 It is therefore not unreasonable to expect a connection between the Wilson loops we have computed and certain large-$N$ Kondo models. 
In this context it would  be interesting to understand the physical origin of the 
non-zero  effective ``temperature'' we see in our zero temperature calculations, both for ${\cal N}=4$ theory \eqref{n=4anti} and for the ${\cal N}=2$ SCFT \eqref{fermion}. In the former case the effective ``temperature'' associated to the fermion impurity action scales as $1/\sqrt{\lambda}$ and parametrizes the stringy corrections to the D5-brane action computing the antisymmetric Wilson loop. For the ${\cal N}=2$ SCFT at infinite 't Hooft coupling the corresponding equation \eqref{fermion} which fixes the occupation number of the impurity fermions has no free parameters and appears to be fixed at an effective temperature of ${\cal O}(1)$. The resulting action for the antisymmetric Wilson loop Eq.\eqref{smallf1} bears a striking resemblance to the impurity entropy for 
the large-$N$ multichannel Kondo model of \cite{parcollet} (setting $K=N$ in that model). In particular, both share the same logarithmic
dependence on the fermion occupation number $k/N$.

Motivated by this resemblance between the two systems, we make a purely empirical observation which appears to follow solely from the strong coupling eigenvalue density $\rho_\infty(x)= 1/(2\cosh\tfrac{\pi x}{2})$. Suppose that we  introduce a fictitious ``temperature'' $\beta$ into our large-$N$ equations for the antisymmetric loop so the equation for the fermion occupation number \eqref{fermion} reads,
\be
\int_{-\infty}^\infty dx\, \frac{1}{2\cosh\frac{\pi x}{2}}\,\frac{1}{1+e^{2\pi\beta(x-z)}}\,=\,f.
\ee
In the zero temperature limit $\beta\to\infty$, where only the states below the chemical potential $z$ are occupied, we obtain that the saddle point (chemical potential) and impurity action are
\be
z=\tfrac{2}{\pi}\,\ln\,\tan\left(\tfrac{\pi f}{2}\right)\,,\quad
%Substituting this into the impurity action, Eq.\eqref%{integral1}, we find 
%\be
\ln \langle W_{A_k}\rangle\big|_{\beta\to\infty}\,=\,
\frac{8 }{\pi}\beta N\left[{\rm Im}\,{\rm Li}_2\left(i \tan\tfrac{\pi f}{2}\right)-\tfrac{f\pi}{2}\ln\,\tan\left(\tfrac{\pi f}{2}\right)\right],
\ee
When expanded out for small $f =\tfrac{k}{N}$ this yields,
\be
\ln \langle W_{A_k}\rangle\big|_{\beta\to\infty}\,=\,
4\beta N\left[1-\ln\left(\frac{\pi}{2}\frac{k}{N}\right)-\frac{\pi^2}{36}\left(\frac{k}{N}\right)^3+\ldots\right]\,,
\ee
 precisely matching the impurity entropy for the large-$N$ multichannel Kondo model \cite{parcollet, hkt}. The physical significance of this is unclear at the moment as we do not really have a means to introduce a tunable effective ``temperature'' in the matrix model for the $N_f=2 N$ theory. In the ${\cal N}=4$ theory, as explained earlier, the 't Hooft coupling appears to play this role. It is possible that studying Wilson  loops in more general ${\cal N}=2$ quiver SCFT's (such as the $SU(N)\times SU(N)$ theory) might shed light on this connection, if any, to Kondo models. 
 
\paragraph{\underline {Aspects of large-$N$ duals}:} The study of supersymmetric Wilson loops in even larger representations, with ranks of order $N^2$, proved to be remarkably fruitful in the context of the ${\cal N}=4$ theory. The relation between the matrix model picture for such operators in terms of back-reacted eigenvalue distributions \cite{yamaguchi1, Okuda:2007kh}, and their gravity duals, has been understood  via explicit construction of half-BPS type IIB ``bubbling geometries''
\cite{lunin, deg}. These geometries
 incorporate the complete back-reaction of the D3 and D5-branes dual to the large Wilson loop operators. It would be interesting to pursue the study of these large rank (${\cal O}(N^2)$) Wilson loops in the $N_f=2N$ theory using Pestun's matrix model. It is possible that the large-$N$  behaviour of these operators will provide us with further clues about the large-$N$ string dual of this theory. The first steps towards this computation are sketched in the Appendix.

Finally, it has been argued recently \cite{gpr} that the string dual to the ${\cal N}=2$ SCFT in the Veneziano limit is a non-critical string background with seven geometric dimensions. It would be extremely interesting to know if the results found for large Wilson loops in this paper could be naturally interpreted as expected features of branes in such non-critical backgrounds.

 \acknowledgments{We would like to thank Tim Hollowood for enjoyable discussions. This research was supported in part by the STFC grant ST/G000506/1.}

\newpage
\appendix
\section{$O(N^2)$ Young Tableaux in $\mathcal{N}=2$ SCQCD}

In this appendix we derive a matrix model for `even larger' representations of $SU(N)$, i.e. those whose Young tableaux are made up of ${\cal O}(N^2)$ boxes, by restating a derivation appearing in \cite{okudahalmagyi}. 

There is a well known formula for the character of an $SU(N)$ representation, expressed in terms of its Young tableau:
\be
\Tr_R e^{2\pi a}=\frac{\mathrm{det}(e^{2\pi (N+R_i-i)a_j})}{\mathrm{det}(e^{2\pi(N-i)a_j})}
\ee
The following formulae hold:
\begin{align}
\mathrm{det}A_{N\times N}&=\sum_{\sigma\in S_N} {\rm sgn}(\sigma) \prod_i A_{i\,\sigma(i)}\\
\mathrm{det}(e^{2\pi(N-i)a_j})&=\prod_{i<j}(e^{2\pi a_i}-e^{2\pi a_j})\,.
\end{align}
We want to insert this into the path integral to find the matrix model for the Wilson loop in representation $R$:
\bea
\braket{W_R} &=& \frac{1}{Z}\int[\mathrm{d}a]\,\cdots\,\Tr_R e^{2\pi a}\\\nonumber 
&=&  \frac{1}{Z}\int[\mathrm{d}a]\,\cdots\,\sum_{\sigma\in S_N} {\rm sgn}(\sigma) \prod_i e^{2\pi(N+R_i-i)a_{\sigma(i)}}\,\left(\prod_{i<j} (e^{2\pi a_i}-e^{2\pi a_j})\right)^{-1}\,.
\eea
Manipulating the sum over the symmetric group:
\be
\frac{1}{Z}\int[\mathrm{d}a]\,\cdots\,\sum_{\sigma\in S_N} \frac{\prod_i e^{2\pi(N+R_i-i)a_{\sigma(i)}}}{\prod_{i<j} (e^{2\pi a_{\sigma(i)}}-e^{2\pi a_{\sigma(j)}})}
\ee
We now use the fact that the $a_i$'s are `dummy variables' within the integral, and so we can relabel them at will. This leads to
\be
\frac{1}{Z}\int[\mathrm{d}a]\,\cdots\,N!\,\frac{\prod_i e^{2\pi (N+R_i-i)a_i}}{\prod_{i<j} (e^{2\pi a_i}-e^{2\pi a_j})}
\ee
Now we label the eigenvalues. We use the notation of e.g. \cite{okudahalmagyi}. The eigenvalues split into groups labelled by the blocks of the Young tableau:
\begin{align*}
&u_1^{(1)} \cdots u_{n_1}^{(1)}\\
&u_1^{(2)} \cdots u_{n_2}^{(2)}\\
&\cdots\\
&\cdots\\
&u_1^{(m)} \cdots u_{n_m}^{(m)}\\
&u_1^{(m+1)} \cdots u_{n_{m+1}}^{(m+1)}
\end{align*}
We then have:
\bea \int \cdots N!\prod_I\prod_i &&e^{2\pi(N-(N_{m+2-I}+i)+K_I)a_i^{(I)}}\,\times\\\nonumber
 &&\times\left(\prod_I\prod_{i<j}(e^{2\pi a_i^{(I)}}-e^{2\pi a_j^{(I)}})\prod_{I<J}\prod_{i,j}(e^{2\pi a_i^{(I)}}-e^{2\pi a_j^{(J)}})\right)^{-1}\eea
Now we can use the identity
\be \prod_{I<J}\prod_{i,j}(e^{2\pi a_i^{(I)}}-e^{2\pi a_j^{(J)}}) = e^{2\pi \sum_I(N-N_{m+1-I})\sum_i a_i^{(I)}} \prod_{I<J}\prod_{i,j}(1-e^{2\pi (a_j^{(J)}-a_i^{(I)})})\ee
to obtain
\be \int \cdots N!\prod_I\prod_i e^{2\pi (n_I-i+K_I)a_i^{(I)}}\,\prod_{I<J}\prod_{i,j}(1-e^{2\pi (a_j^{(J)}-a_i^{(I)})})^{-1} \prod_I\prod_{i<j}(e^{2\pi a_i^{(I)}}-e^{2\pi a_j^{(I)}})^{-1}\ee
We can again manipulate the dummy `$a$' variables to write
\be \int [d a] \cdots N!\prod_I\prod_ie^{K_I 2\pi a_i^{(I)}}\,\sum_{\sigma\in S_N} \frac{\prod_i \
e^{2\pi(n_I-i)a_{\sigma(i)}^I}}{\prod_{i<j} (e^{2\pi a_{\sigma(i)}^I}-e^{2\pi a_{\sigma(j)}^I})} \prod_{I<J}\prod_{i,j}(1-e^{2\pi a_j^{(J)}-2\pi a_i^{(I)}})^{-1}\ee
But using the identity for the determinant we stated earlier, the sum in the second term here is equal to one. Including the Vandermonde determinant, we can write our final result for the matrix model partition function as
\bea 
\int \left(\prod_{I,i} da_i^{(I)} \right) \cdots N!\prod_I &&\left(\prod_i e^{K_I 2\pi a_i^{(I)}}\prod_{i<j} (a_i^{(I)}-a_j^{(I)})^2
\right)\,\times\\\nonumber
&&\times \prod_{I<J} \left(\frac{\prod_{i,j}(a_i^{(I)}-a_j^{(J)})^2}{\prod_{i,j}(1-e^{2\pi a_j^{(J)}-2\pi a_i^{(I)}})} \right)\,.
\eea
The new terms in this expression introduced by the trace insertion are those involving exponentials: there is an extra force on each block of eigenvalues (labelled by $I$) proportional to $K_I$, and there is an inter-block interaction. 

So we write down the matrix model to be solved to evaluate large tableaux in the $\mathcal{N}=2$ theory:
\bea
\braket{W_R}=&&\frac{1}{\mathcal{Z}} \int \left(\prod_I \prod_i da_i^{(I)} 
e^{-\frac{8\pi^2 N}{\lambda}(a_i^{(I)})^2 + 2\pi K_I a_i^{(I)}} H^{2N}(a_i^{(I)}) \prod_{i<j} \frac{(a_i^{(I)}-a_j^{(I)})^2}{H^2(a_i^{(I)}-a_j^{(I)})}
\right)
\nonumber\\
 &&\times \prod_{I<J} \prod_{i,j}\left(\frac{(a_i^{(I)}-a_j^{(J)})^2}{(1-e^{a_j^{(J)}-a_i^{(I)}}) H^2(a_i^{(I)}-a_j^{(J)})} \right)\,.
\eea

This has saddle point equations:

\bea
 &&\frac{8\pi^2}{\lambda} a_i^{(I)} - 2\pi \frac{K_I}{2N} - K(a_i^{(I)}) = \frac{1}{N}\sum_{i(\neq j)} \left( \frac{1}{a_i^{(I)}-a_j^{(I)}} - K(a_i^{(I)}-a_j^{(I)})\right)\\\nonumber
 &&+ \frac{1}{N} \sum_{J(\neq I)}\sum_j \left(\frac{1}{a_i^{(I)}-a_j^{(J)}} - K(a_i^{(I)}-a_j^{(J)}) - \frac{1}{2}\frac{1}{1-e^{2\pi a_i^{(I)}-2\pi a_j^{(J)}}} \right)\,.
\eea

If the model behaves qualitatively like that for $\mathcal{N}=4$ SYM, the different groups of eigenvalues will be situated at intervals along the real line, with the $a^{(1)}_i$  furthest to the right, and each group having a profile as for the model without the trace insertion. This picture should be correct if one could establish that the eigenvalue spread within each group ($\sim\log\lambda$) were parametrically larger than the spacings between groups. This would be an interesting avenue for future work. For $\mathcal{N}=4$ this model yields a prediction for the IIB supergravity action evaluated on a weakly curved `bubbling geometry'. The equivalent here would be informative about the curvatures involved in a gravity dual. 

%\startappendix
%\Appendix{}
%\label{sec:appendix}

\end{document}